\documentclass[final,journal,letterpaper]{IEEEtran}

 \usepackage[hidelinks]{hyperref}
 \usepackage{varioref} %smart page, figure, table, and equation referencing
 \usepackage{wrapfig} %wrap figures/tables in text (i.e., Di Vinci style)
 \usepackage{threeparttable} %tables with footnotes
 \usepackage{dcolumn} %decimal-aligned tabular math columns
  \newcolumntype{d}{D{.}{.}{-1}}
 \usepackage{nomencl} %nomenclature generation via makeindex
  \makeglossary
 \usepackage{subfigmat} % matrices of similar subfigures, aka small mulitples
 \usepackage{fancyvrb} %extended verbatim environments
  \fvset{fontsize=\footnotesize,xleftmargin=2em}
 \usepackage{lettrine} %dropped capital letter at beginning of paragraph

 \usepackage{epsfig}
 \usepackage{epstopdf}
 \usepackage{subfigure}
 \usepackage{latexsym}
 \usepackage{amsmath}
 \usepackage{amsfonts}
 \usepackage{amssymb}
 \usepackage{mathrsfs}
 \usepackage{cases}
 \usepackage{array}
 \usepackage{color}
 \usepackage{soul}
 \setulcolor{red}
 \sethlcolor{yellow}
 \usepackage{url}
 \usepackage[noadjust]{cite}
 \usepackage{graphicx}
 \usepackage{psfrag}
 \usepackage{color}
 \usepackage[dvipsnames]{xcolor}
 \usepackage{multirow}
 \usepackage{ragged2e}
 \usepackage{graphicx}  
 \usepackage{makecell}
 \usepackage{tabularx}%for autmatic line break in table cell

 \newcommand*{\Scale}[2][4]{\scalebox{#1}{$#2$}}%
 \usepackage{breqn}
 \newcommand{\comment}[1]{}
 \usepackage{algorithm} 
\usepackage{algpseudocode} 
 \DeclareMathOperator{\argmax}{argmax}
\begin{document}
%
% paper title
% Titles are generally capitalized except for words such as a, an, and, as,
% at, but, by, for, in, nor, of, on, or, the, to and up, which are usually
% not capitalized unless they are the first or last word of the title.
% Linebreaks \\ can be used within to get better formatting as desired.
% Do not put math or special symbols in the title.
\title{{Efficient Learning of Voltage Control Strategies via Model-based Deep Reinforcement Learning}

%
%
% author names and IEEE memberships
% note positions of commas and nonbreaking spaces ( ~ ) LaTeX will not break
% a structure at a ~ so this keeps an author's name from being broken across
% two lines.
% use \thanks{} to gain access to the first footnote area
% a separate \thanks must be used for each paragraph as LaTeX2e's \thanks
% was not built to handle multiple paragraphs
%

\author{Ramij R. Hossain,~\IEEEmembership{Student Member,~IEEE}, Tianzhixi Yin~\IEEEmembership{Member,~IEEE}, Yan Du, Renke Huang,~\IEEEmembership{Senior Member,~IEEE,}
          Jie Tan,  Wenhao Yu,  Yuan Liu,~\IEEEmembership{Member,~IEEE}, Qiuhua Huang,~\IEEEmembership{Member,~IEEE}}
%\vspace{-0.3in}}

        % <-this % stops a space
\thanks{The project is supported by funding from the U.S. Department of Energy (DOE) Advanced Research Projects Agency - Energy (ARPA-E) OPEN 2018 program. (\textit{Corresponding author: Tianzhixi Yin, Ramij R. Hossain})}
\thanks{Tianzhixi Yin, Yuan Liu are with Pacific Northwest National Laboratory (PNNL), Richland, WA 99354, USA (e-mail: \{tianzhixi.yin, yuan.liu\}@pnnl.gov).}
\thanks{Qiuhua Huang, Renke Huang and Yan Du were with PNNL. (e-mail:  qiuhua.huang@ieee.org, huangrenke@gmail.com, ydu0116@gmail.com). Ramij R. Hossain was a student intern at PNNL while contributing to this work (e-mail: rhossain@iastate.edu).}
\thanks{J. Tan and W. Yu are with Google Brain, Google Inc, Mountain View, CA, 94043 USA (e-mail: \{jietan, magicmelon\}@google.com).} 
 }

% note the % following the last \IEEEmembership and also \thanks - 
% these prevent an unwanted space from occurring between the last author name
% and the end of the author line. i.e., if you had this:
% 
% \author{....lastname \thanks{...} \thanks{...} }
%                     ^------------^------------^----Do not want these spaces!
%
% a space would be appended to the last name and could cause every name on that
% line to be shifted left slightly. This is one of those "LaTeX things". For
% instance, "\textbf{A} \textbf{B}" will typeset as "A B" not "AB". To get
% "AB" then you have to do: "\textbf{A}\textbf{B}"
% \thanks is no different in this regard, so shield the last } of each \thanks
% that ends a line with a % and do not let a space in before the next \thanks.
% Spaces after \IEEEmembership other than the last one are OK (and needed) as
% you are supposed to have spaces between the names. For what it is worth,
% this is a minor point as most people would not even notice if the said evil
% space somehow managed to creep in.

% The paper headers
%\markboth{Journal of \LaTeX\ Class Files,~Vol.~14, No.~8, August~2015}%
%{Shell \MakeLowercase{\textit{et al.}}: Bare Demo of IEEEtran.cls for IEEE Journals}
% The only time the second header will appear is for the odd numbered pages
% after the title page when using the twoside option.
% 
% *** Note that you probably will NOT want to include the author's ***
% *** name in the headers of peer review papers.                   ***
% You can use \ifCLASSOPTIONpeerreview for conditional compilation here if
% you desire.

% make the title area
\maketitle
% {\color{blue}This is just a draft version, needs to be changed a lot. I am just structuring the paper before starting section III. (Ramij)}

% As a general rule, do not put math, special symbols or citations
% in the abstract or keywords.

\begin{abstract}
This article proposes a model-based deep reinforcement learning (DRL) method to design emergency control strategies for short-term voltage stability problems in power systems. Recent advances show promising results in model-free DRL-based methods for power systems, but model-free methods suffer from poor sample efficiency and training time, both critical for making state-of-the-art DRL algorithms practically applicable. DRL-agent learns an optimal policy via a trial-and-error method while interacting with the real-world environment. And it is desirable to minimize the direct interaction of the DRL agent with the real-world power grid due to its safety-critical nature. Additionally, state-of-the-art DRL-based policies are mostly trained using a physics-based grid simulator where dynamic simulation is computationally intensive, lowering the training efficiency. We propose a novel model-based-DRL framework where a deep neural network (DNN)-based dynamic surrogate model, instead of a real-world power-grid or physics-based simulation, is utilized with the policy learning framework, making the process faster and sample efficient. However, stabilizing model-based DRL is challenging because of the complex system dynamics of large-scale power systems. We solved these issues by incorporating imitation learning to have a warm start in policy learning, reward-shaping, and multi-step surrogate loss. Finally, we achieved 97.5\% sample efficiency and 87.7\% training efficiency for an application to the IEEE 300-bus test system.

\end{abstract}

% Note that keywords are not normally used for peerreview papers.
\begin{IEEEkeywords}
Deep reinforcement learning, Voltage stability, Model-based deep reinforcement learning, Imitation learning, Augmented random search
\end{IEEEkeywords}

\section{Introduction}
\IEEEPARstart{E}{mergency} control design in power and energy systems is becoming significantly important owing to the rapid transformation of electric power landscapes. The proliferation of distributed energy resources (DERs) and dynamic loads creates operational challenges and makes power systems vulnerable. Unless otherwise tackled appropriately, severe disturbances can destabilize the system and create large-scale blackouts \cite{aemo}. At this juncture, recent studies 
\cite{glavic2019deep,cao2020reinforcement,chen2022reinforcement} show that deep reinforcement learning (DRL) methods can provide more adaptive yet faster solutions compared to the traditional rule-based and optimization-based (e.g., model predictive control) methods in power system stability and emergency control applications. DRL-based methods in power systems typically learn optimal policies in a model-free manner \cite{yan2018data,huang2019adaptive,yan2020multi,duan2019deep,kamruzzaman2021deep}, which is an appealing virtue considering high complexity of power systems. However, owing to direct interaction with the environment during policy learning, model-free methods are not sample efficient \cite{luo2022survey}, thus unsuitable for direct application in real-world power grids, where trial-and-error effects are highly costly. In the literature, DRL-based control policies are predominantly trained in a simulated environment instead of directly interacting with the real-world power grid. For first-principle-based dynamic modeling of bulk power system systems, large-scale differential algebraic equation (DAE) models have to be solved, which is computationally intensive. Considering the large amount of explorations required for DRL training, this leads to extensive training and tuning time to obtain a good DRL-based control policy for large-scale power systems. Therefore, the current state-of-the-art model-free DRL algorithms in power systems \cite{yan2018data,huang2019adaptive,yan2020multi,duan2019deep,kamruzzaman2021deep} suffer the major challenges related to (a) sample complexity, and (b) training time.

 Model-based DRL (MB-DRL) has been shown to be more sample efficient compared to model-free methods \cite{wang2019benchmarking}. Unlike model-free approaches, model-based approaches learn a surrogate (or transition) model of the system dynamics and obtain an optimal policy through interaction with the learned surrogate model. While there are some recent works in applying them in power system steady-state and quasi-steady-state applications \cite{cao2022model,shuai2020online,kamel2021data}, there are some grand challenges in applying MB-DRL in power system dynamic stability control such as deriving a sufficiently accurate dynamic model for large, complex systems, modeling error accumulation, and state-action distribution drifting. To the best of our knowledge, \textit{there is no systematic study of MB-DRL methods in terms of feasibility and applicability for bulk power system dynamic stability control applications}. In this paper, we developed a novel MB-DRL framework for bulk power system voltage stability control that significantly accelerates the training process and improves sample efficiency compared to the state-of-the-art model-free methods. We overcame the challenges mentioned above in training DRL policy with a learned model by introducing (i) multi-step loss in model learning, (ii) adaptive model update, (iii) robust reward structure, and (iv) imitation learning. Our method resulted in 97.5\% and 87.7\% reductions in sample complexity and training time, respectively. We believe this is a major boost in making DRL practical for real-world grid dynamic control applications.

\begin{table*}[t]
\centering
\tabcolsep=0.055 cm
\caption{Literature Summary on Model-free RL/DRL in Emergency Control of Power Systems}
\label{tab:litreview1}
\begin{tabular}{|c|c|c|c|c|c|}
\hline
\textbf{References}          & \textbf{Application}              & \textbf{Algorithm}   & \textbf{Control Action} & \textbf{Model-free/Model-based} & \textbf{Surrogate Model} \\ \hline
\cite{huang2019adaptive}    & Emergency voltage control   & DQN                   & Load-shedding & Model-free & $\times$ \\ \hline
\cite{li2022supervised}     & Emergency voltage control   & Dueling DDQN   + Behavior Cloning   & Load-shedding & Model-free & $\times$ \\ \hline
\cite{huang2021accelerated} & Emergency voltage control   & ARS                   & Load-shedding & Model-free & $\times$ \\ \hline
\cite{huang2022learning}    & Emergency voltage control   & ARS + Meta   Learning & Load-shedding & Model-free & $\times$ \\ \hline
\cite{jiang2019power}       & Emergency voltage control   & PPO                   & Load-shedding & Model-free & $\times$ \\ \hline
\cite{li2021research}       & Emergency voltage control   & DDPG and DQN          & Load-shedding & Model-free & $\times$ \\ \hline
\cite{hossain2021graph}     & Emergency voltage control   & Graph Convolutional Network + DDQN            & Load-shedding & Model-free & $\times$ \\ \hline
\cite{zhang2018load} & Emergency voltage control & DDPG & Load shedding & Model-free & $\times$ \\ \hline
\cite{yan2020multi} & Emergency frequency control & DDPG & Load shedding & Model-free & $\times$ \\ \hline
\cite{xie2021distributional} & Emergency frequency control & Distributional   SAC & Load-shedding           & Model-free & $\times$    \\\hline
\cite{chen2020model} & Emergency frequency control & Multi-Q-learning+DDPG & Load shedding & Model-free & $\times$ \\ \hline
\end{tabular}
\end{table*}
\begin{table*}[t]
\centering
\tabcolsep=0.28 cm
\caption{Literature Summary on Model-based RL/DRL and application of Surrogate models in Power Systems}
\label{tab:litreview2}
\begin{tabular}{|c|c|c|c|c|}
\hline
\textbf{References} &
  \textbf{Reinforcement Learning} &
  \textbf{Surrogate Model} &
  \textbf{Emergency Control} &
  \textbf{Application} \\ \hline
\cite{WANG2021116722}     & $\checkmark$ & $\checkmark$ & $\times$ & Retail energy pricing                          \\ \hline
\cite{cao2022model}       & $\checkmark$ & $\checkmark$ & $\times$ & Voltage stabilization (Power flow-based study) \\ \hline
\cite{shuai2020online}    & $\checkmark$ & $\checkmark$ & $\times$ & Optimal scheduling of residential micro-grids  \\ \hline
\cite{kamel2021data}      & $\checkmark$ & $\times$  & $\times$ & Branch Stress Reduction                        \\ \hline
\cite{gao2022model}       & $\checkmark$ & $\checkmark$ & $\times$ & Volt-var control (Power-flow based)            \\ \hline
% \cite{dalal2019chance} & $\times$ & $\checkmark$ & $\times$ & Planning for outage scenarios \\ \hline
% \cite{balduin2019towards} & $\times$  & $\checkmark$ & $\times$ & Simulation of low voltage power grid           \\ \hline
% \cite{qiu2020analytic}    & $\times$  & $\checkmark$ & $\times$ & Transfer capability operational planning       \\ \hline
% \cite{ren2021}    & $\times$  & $\checkmark$ & $\times$ & Transient stability assessment       \\ \hline
\end{tabular}
\end{table*}
\subsection{Literature review}
Learning a policy from scratch without prior knowledge of the underlying process is daunting. Model-based RL (MBRL) is a suitable alternative to facilitate this process. Model-based planning and learning has been discussed extensively in RL literature \cite{sutton2018reinforcement,schaal1997learning,schneider1997exploiting,atkeson1997comparison}. A comprehensive survey on MBRL methods can be found in the recent review \cite{luo2022survey}. The application area of standard approaches on MBRL is mostly focused on games, robotic control, and autonomous driving. We mainly focus on past works applying MBRL methods in power systems, but before discussing MBRL methods, for sake of completeness, we present a brief review on model-free RL works in power systems.

\subsubsection{Model-free RL in Power Systems}
A significant number of previous efforts utilizing model-free RL methods for power systems applications can be found in a recent review \cite{chen2022reinforcement}. The domain of applications broadly comprises frequency regulation, voltage control, and energy management. Leaving aside DRL applications utilizing steady-state (or power flow)-based analysis of power systems, we primarily review relevant studies that discuss model-free DRL applications in emergency control. To this end, there are two types of problems, (a) voltage control and (b) frequency control. In emergency voltage control problems, a deep Q network (DQN)-based load shedding strategy is proposed in \cite{huang2019adaptive}. The deep deterministic policy gradient (DDPG) and proximal policy gradient (PPO) methods are utilized in \cite{zhang2018load,jiang2019power}, respectively, for emergency voltage control. Load-shedding for voltage control is also designed by combining dueling double DQN and behavior cloning (BC) methods \cite{li2022supervised}. In earlier works, we developed parallel augmented random search (PARS) \cite{huang2021accelerated}, and deep meta reinforcement learning (DMRL)-based approaches \cite{huang2022learning} to address the issues of faster convergence (training), scalability, and adaptation to new scenarios. The issues of network topology changes are tackled using graph convolutional network (GCN)-based double DQN algorithm in \cite{hossain2021graph}. In frequency control, DDPG is used in \cite {zhang2018load}, while \cite{xie2021distributional} used a novel distributional soft actor-critic (SAC) method. Multi-agent RL with DDPG is explored in \cite{chen2020model} for emergency frequency control. For ease of understanding, we summarized these works in Table~\ref{tab:litreview1}, and it is important to note that these works followed learned optimal policies interacting directly with the grid simulator without using any learned dynamics.

% As our primary objective is to design emergency control, we provided a list of relevant works in Table-I, leaving aside DRL applications utilizing steady-state (or power flow)-based analysis of power systems. 

\subsubsection{Model-based RL and Application of Surrogate Model in Power Systems}
Model-based RL or DRL studies in power systems have been growing recently. Authors in \cite{WANG2021116722} integrated a deep belief network (DBN)-based surrogate model into a DRL framework to optimally select the retail energy prices for the community agents. Reference \cite{cao2022model} designed a surrogate model to approximate the nonlinear mapping from the bus active and reactive power injections to the voltage magnitude connected to the DRL-based control design for voltage stabilization. A model-based DRL algorithm with Monte-Carlo tree search for optimal scheduling of a residential micro-grid is developed in \cite{shuai2020online}. Reference \cite{kamel2021data} introduces a hybrid data-driven and model-based RL for stress reduction of power systems branches. The model-augmented actor-critic method with safety constraints for volt-var control (VVC) is presented in \cite{gao2022model}. This paper learns the environment model for VVC operation utilizing a bootstrap ensemble of probabilistic neural networks. Table~\ref{tab:litreview2} provides a summary of these applications.
% \subsubsection{Surrogate Model in Power Systems}
% Power system dynamic simulations are computationally intensive. Therefore, developing a surrogate model to replicate the dynamics is a time-efficient solution and has been leveraged in a few non-DRL applications in recent years. Reference \cite{dalal2019chance} utilized proxy machine learning model in planning problem for evaluating different outage scenarios for long simulation horizon (for instance, 1 year). The authors assessed that the problem is impractical in terms of computational time using a conventional simulator. Reference \cite{balduin2019towards}, a DNN-based surrogate model is introduced to replicate the operation of a low voltage power grid. In \cite{qiu2020analytic}, a deep-learning-based surrogate model is developed to drastically reduce the real-time computation time of transient stability constrained total transfer capability (TTC) operational planning problem. Reference \cite{ren2021} utilizes surrogate model for transient stability assessment with decision tree based interpretability analysis.

With the above summary of the model-free and model-base DRL-based approaches, we have found the following limitations and research gaps:
% \subsection{Limitations of existing works}
% We identified the following limitations and research gaps:
\begin{itemize}
    % \item The application area of \emph{Standard Approaches on Model-based RL} works is focused particularly on games, robotic control, and autonomous driving.
    \item Existing MBRL (or MB-DRL) studies are based on steady-state formulation and do not consider any power system dynamics. The feasibility and applicability of MB-DRL for emergency control problems with bulk power system dynamics have yet to be addressed.
    \item Model-free methods have shown the potential of RL-based approaches in power systems, but these methods require direct interaction with a grid simulator, therefore, perform poorly in terms of sample complexity and training efficiency.
\end{itemize}

\subsection{Main Contributions}
% The main contribution of this work can be summarized as follows:
To address the above-mentioned issues, 
\begin{itemize}
    \item We developed a novel model-based DRL algorithm for emergency voltage control, MB-PARS, which (a) learns a DNN-based surrogate model to simulate power system dynamics, and (b) utilizes a learned surrogate model to train a DRL agent. In the training of the DRL agent, we utilized fast, adaptive, and derivative-free DRL algorithm PARS \cite{huang2021accelerated}. Moreover, we bring the idea of imitation learning \cite{10.1145/3054912} to provide a warm start to policy learning. The introduction of MB-PARS greatly reduces the DRL training time and sample complexity. 
    % To the best of our knowledge, this is the first model-based DRL algorithm for bulk power system control with transient stability dynamics of power systems being fully considered. 
    \item We incorporated (i) multi-step prediction loss to improve the prediction capability of the surrogate model and (ii) an online update of the surrogate model (in the training phase) to tackle state-action distribution drifting, and (iii) reward shaping to accommodate prediction error during the DRL training to help stabilize the training of MB-PARS method. To the best of our knowledge, this is the first model-based DRL algorithm for bulk power system control with transient stability dynamics of power systems being fully considered. 
\end{itemize}

% combining surrogate model for power system dynamics, imitation learning \cite{10.1145/3054912}, and PARS \cite{huang2021accelerated} for training data-driven power system dynamic voltage control strategies. The introduction of MB-PARS greatly reduces the DRL training time and sample complexity. To the best of our knowledge, this is the first model-based DRL algorithm for bulk power system control with transient stability dynamics of power systems being fully considered. 
% We incorporated (i) utilizing multi-step prediction loss to improve the prediction capability of the surrogate model and (ii) updating the surrogate model in the training phase to tackle state-action distribution drifting, and (iii) reward shaping to accommodate prediction error during the DRL training to help stabilize the training of MB-PARS method. We also leveraged imitation learning before the start of policy training to provide a warm-start in the policy training phase.

With these contributions, the MB-PARS method is applied for determining an optimal load-shedding strategy against the fault induced delayed voltage recovery (FIDVR) problem \cite{potamianakis2006short} in IEEE 300 Bus system. Our study shows that policy training time and sample complexity with MB-PARS can be reduced by 87.7\% and 97.5\%, respectively, compared to its model-free counterpart. In testing phase, for new operating conditions including unseen power flow cases, and new contingencies the performance of the trained policy with surrogate model is proved satisfactory compared to one state-of-art model-free DRL method.

\subsection{Organization}
The rest of the paper is organized as follows: Section \ref{sec2} includes a comprehensive study on model-based reinforcement learning methods. Section \ref{sec3} presents the details of our proposed MB-PARS algorithm  and introduces the voltage control problem considered in this paper. Next, we show the test cases, training and testing results in Section \ref{sec4}. At last, conclusions and future directions are discussed in Section \ref{sec:conclusions}.

\section{Model Based DRL} \label{sec2}

In this section, we provide a brief overview of model-based DRL, its challenges and different components.
\subsection{Overview and Main Challenges}\label{sec2a}
RL is a sequential decision making process, where an agent interacts with an unknown environment (from agent's point of view) and collects some abstract signal known as reward. The problem is studied in a (partial observable) Markov Decision Process (MDP) setting defined by a tuple $(\mathcal{S},\mathcal{A},\mathcal{P},\mathcal{R})$ \cite{sutton2018reinforcement}, where, $\mathcal{S} :=$ state space, $\mathcal{A} :=$ action space and $\mathcal{P}:\mathcal{S}\times\mathcal{A}\rightarrow\mathcal{S}$ is the transition function giving the next state $s_{t+1}\in \mathcal{S}$ for a given current state $s_t\in \mathcal{S}$ and action $a_t\in \mathcal{A}$. Besides, for each state-action pair, the environment returns a reward $r:\mathcal{S}\times\mathcal{A}\times\mathcal{S}\rightarrow\mathcal{R}$. The goal of RL is to learn a policy $\pi: \mathcal{S}\rightarrow\mathcal{A}$ for the agent, such that it maximizes the expected cumulative reward over a horizon $T$, for a given initial state distribution $\rho_d$ and following the transition dynamics given by $\mathcal{P}$. Mathematically, $\pi^{\ast} = \argmax_{\pi} J(\pi)$.
%\begin{gather}
   %\pi^{\ast} = \argmax_{\pi} J(\pi)
%\end{gather}
where, $J(\pi) = \mathbf{E}_{{s_0 \sim \rho_d}} \Big[\sum^{T}_{t=0}{\gamma^t r(s_t,a_t,s_{t+1})}\Big]$, $a_t = \pi(s_t) $, $s_{t+1}$ is given by $\mathcal{P}$, and $\gamma$ is the discount factor. In DRL set-up, the policy is parameterized by $\theta$, hence the cumulative reward function $J(\pi)$ becomes $J(\pi_{\theta}) = \mathbf{E}_{{s_0 \sim \rho_d}} \Big[\sum^{T}_{t=0}{\gamma^t r(s_t,\pi(\theta,s_t)}\Big]$, an implicit function of $\theta$, and consequently, the problem converts to find the $\theta^{\ast}$, which maximizes $J(\pi_{\theta})$, i.e., $\theta^{\ast} = \argmax_{\theta} J(\pi_{\theta})$. 

In model-based RL, the transition dynamics $\mathcal{P}$ can be represented by, (a) a stochastic function $P[s_{t+1}|s_t,a_t]$, which can be modeled as a Gaussian process \cite{pilco}, $ P[s_{t+1}|s_t,a_t] = \mathcal{N}(s_t|\mu_t,\sigma_t)$, where, $\mu_t$ and $\sigma_t$ are prior mean and standard deviation, or (b) a deterministic function, parameterized by $\phi$, with dynamics $s_{t+1} = f_{\phi}(s_t,a_t)$ \cite{pmlr-v100-yang20a}. In both cases, the model representing the transition dynamics must be learnt from prior collected trajectory data. Now, the learned model is utilized to estimate the next-state $\hat{s}_{t+1}$, for a particular state-action pair $\{s_t,a_t\}$. This estimated next state $\hat{s}_{t+1}$ paired with $\{s_t,a_t\}$, are used to compute the reward $r_t$ and thereby the cumulative discounted reward over the entire episode starting for $t=0$ to $T$. Thus, the accuracy of the cumulative reward computation depends solely on the accuracy of the probabilistic or parameterized model under consideration, and importantly, the cumulative reward is optimized to retrieve the optimal policy. This leaves the major challenge in model-based RL compared to model-free RL which does not need any model to learn and can directly interact with the environment. 
%\subsection{Challenges}

\subsection{DNN-based Modeling of Transition Dynamics}\label{sec2b}
In this work, we modeled the power systems transition dynamics using a DNN $f_{\phi}(s_t,a_t)$, where $\phi$ represents the parameter (weights and bias vectors) of the DNN. The most common way is to directly map the next state $\hat{s}_{t+1}$ as a function of current state $s_t$, and action $a_t$, implying $s_{t+1}$ as the output of  $f_{\phi}(\cdot,\cdot)$ while $s_t$ and $a_t$ are the inputs. The data to train $f_{\phi}(\cdot,\cdot)$ can be collected executing a random policy for different initial state $s_0 \sim \rho_d$ on the actual environment for the horizon $t=0$ to $T$, resulting in a collection of ground-truth trajectory data of the form, $\tau = \{(s_0,a_0),(s_1,a_1),\cdots,(s_{T},a_{T})\}$. For a given set of state-transition tuple $\mathcal{D} = \{(s_t,a_t,s_{t+1})\}$, $f_{\phi}(\cdot,\cdot)$ can be trained by minimizing $    \mathcal{L}(\phi) = \frac{1}{\lvert\mathcal{D}\rvert} \sum_{(s_t,a_t,s_{t+1})\in \mathcal{D}}\lVert {s_{t+1} - f_{\phi}(s_t,a_t)}\rVert_2^2$. 

But, the current structure of $f_{\phi}(s_t,a_t)$ is not efficient, as mentioned in \cite{nagabandi2018neural}, when $s_t$ and $s_{t+1}$ are very similar. Therefore, the learned model $f_{\phi}(s_t,a_t)$ is more prone to make mistakes in predicting future states. Moreover, the problem aggravates with (a) complex dynamics (for instance in power systems) and (b) small duration between two consecutive time step. Additionally, in DRL context, we need to to propagate the dynamics for long horizon roll-outs, where inaccurate dynamics causes the accumulation of error over the entire horizon. Therefore, to circumvent these issues, we utilized (a) the difference between two consecutive state in modeling $f_{\phi}(\cdot,\cdot)$, where, $s_{t+1} - s_t = f_{\phi}(s_t,a_t)$, and (b) multi-step loss function (\ref{eqn_mult_step_loss}) in contrast to the single-step loss function $    \mathcal{L}_{ss}(\phi) = \frac{1}{\lvert\mathcal{D}\rvert} \sum_{(s_t,a_t,s_{t+1})\in \mathcal{D}}\lVert {s_{t+1} - s_t - f_{\phi}(s_t,a_t)}\rVert_2^2$ while learning $f_{\phi}(s_t,a_t)$.

Following steps are utilized to formulate the multi-step ($M$-step) loss based training of $f_{\phi}(\cdot,\cdot)$:

\begin{itemize}
    \item[(a)] Make the $M$-step ground-truth transition tuple, $\mathcal{D}_{M} = \{(s_{t},a_{t},s_{t+1},a_{t+1},\cdots,s_{t+M})\}$
    \item[(b)] Predict the future states starting from $s_t$, using $\hat{s}_{t+\tau+1} = \hat{s}_{t+\tau} + f_{\phi}(\hat{s}_{t+\tau},a_{t+\tau})$, for $\tau=0$ to $M$. It is important to note that except the first step the predicted states $\hat{s}_{t+\tau}$, instead of the ground true state ${s}_{t+\tau}$, is used as the input of the DNN $f_{\phi}(\cdot,\cdot)$. The usage of the predicted state as the input of the DNN, helps to mitigate the error accumulation for long horizon prediction. 
    \item[(c)] The DNN is trained using mini-batch stochastic gradient descent minimizing the loss given by (\ref{eqn_mult_step_loss}).
\end{itemize}
\vspace{-0.1in}
\small
\begin{multline}\label{eqn_mult_step_loss}
    \mathcal{L}_{ms}(\phi) = \frac{1}{\lvert\mathcal{D}_M\rvert\times M} \sum_{\substack{(s_{t:t+M},\\a_{t:t+M-1})\\\in \mathcal{D}_{M}}} \sum_{\tau=0}^{M-1} \Big \lVert {s_{t+\tau+1} - \hat{s}_{t+\tau+1}}\Big \rVert_2^2
\end{multline}
\normalsize

\subsection{Policy Derivation with the Learned Dynamics and its implications}\label{sec2c}
After learning the model dynamics, the next important step is to learn policy using learned model. For this work, we adopted derivative free PARS, developed in \cite{huang2021accelerated}. Most of the state-of-the-art gradient-based and value-based algorithms are hard to manage in large-scale power grid problems. The details can be found in \cite{huang2021accelerated}. In short, the reasons are (a) ineffective action-space exploration, (b) difficulties in gradient computation due to non-smoothness in the environment and reward structure, (c) challenges in parallel implementation, and (d) high sensitivity to the hyper-parameters, which makes the training notoriously difficult. Unlike gradient-based and value-based methods, the derivative-free techniques are (a) easy to implement, (b) ease of parallelism, and (c) easy to tune (due to lesser number of hyper-parameters).

Next we will consider the following implications, that, in general, model-based DRL faces:
\begin{itemize}
    \item The optimality of the learned policy is closely reliant on the accuracy of the trained dynamical model. Even after taking necessary steps, for instances, multi-step loss, difference of states in the learning of $f_{\phi}(s_t,a_t)$, the performance of learned policy falls drastically if the learned dynamic model falls in an unknown area of state-action space (not observed in the training phase of $f_{\phi}(s_t,a_t)$) during policy training. 
    \item Creation of a comprehensive data-set (for the training of $f_{\phi}(s_t,a_t)$) representing the entire state-action space is challenging and impossible in the case of high dimensional problem setting, like power systems. 
\end{itemize}

To solve these issues, we need to reduce the mismatch between state-action distribution used in the training of $f_{\phi}(s_t,a_t)$ and the state-action distribution observed under current policy $\pi_{\theta}(\cdot)$ \cite{ross2011reduction}. Note that, during training the policy function $\pi_{\theta}(\cdot)$ changes as the learning progresses. This is achieved by (a) collecting new ground-truth transition data using the current policy $\pi_{\theta}(\cdot)$, and add it to the initial transition data collected at the beginning of training surrogate model $f_{\phi}(s_t,a_t)$, and (b) retraining $f_{\phi}(s_t,a_t)$ in the regular interval in the policy learning framework, with a mix of old and new data.

\section{MB-PARS Framework for Voltage Control} \label{sec3}
This section presents the main algorithms and the architecture details of our proposed MB-PARS method for FIDVR related voltage stabilization problem. 
\subsection{Parallel Augmented Random Search}
Parallel Augmented Random Search (PARS), developed in our previous work \cite{huang2021accelerated}, is a scaled up version of ARS algorithm \cite{mania2018simple} to tackle large-scale grid control problems. PARS is a nest parallelism scheme utilizing the inherent parallelism found in ARS, and is implemented using the Ray framework \cite{moritz2018ray}. As mentioned earlier, the main objective in policy search is to find $\theta$ which maximizes (\ref{exp_rew}).
\begin{align}\label{exp_rew}
   J(\pi_{\theta}) = \mathbf{E}_{{s_0 \sim \rho_d}} \Big[\sum^{\mathbf{T}}_{t=0}{\gamma^t r(s_t,\pi(\theta,s_t)}\Big]  
\end{align}
\noindent In (\ref{exp_rew}), ${s_0 \sim \rho_d}$ represents the randomness of the environment and can be encoded by $\xi := {s_0 \sim \rho_d}$, and this simplifies the expression in (\ref{exp_rew}) into $J(\pi_{\theta}) = \mathbf{E}_{\xi} [\mathbf{r}(\pi_{\theta},\xi)]$, where, $\mathbf{r}(\pi_{\theta},\cdot) = \sum^{\mathbf{T}}_{t=0}{\gamma^t r(s_t,\pi(\theta,s_t)}$ is the reward achieved by the policy $\pi_{\theta} = \pi(\theta,\cdot)$  for a single trajectory rollout. Unlike the existing policy gradient algorithms (PPO, SAC, DDPG, TRPO, TD3, A3C), PARS performs direct exploration in parameter $\theta$ space rather than the action space. The detailed explanation of this algorithm can be found in \cite{mania2018simple,huang2021accelerated}. In short, PARS is as follows: i) the algorithm selects random noises $\delta_1,\cdots,\delta_N$ to perturb the policy parameter $\theta$, ii) Utilize the perturbed direction, generates rollouts and computes episode rewards $\mathbf{r}(\pi_{\theta},\cdot)$, iii) Finally, selects the top-performing directions and updates $\theta$ making it align in the best possible direction. The algorithm uses a finite difference approximation to modify $\theta$ instead of back-propagation. It consequently makes the algorithm easy-to-tune with a lesser number of hyper-parameters than policy gradient algorithms. 

\subsection{MB-PARS algorithm and Implementation Architecture}
We propose a novel MB-PARS algorithm that combines the concept of model-based RL detailed in Section \ref{sec2} and the PARS algorithm to accelerate the control policy learning in large-scale grid control problems. 
% As reported in \cite{huang2021accelerated}, undoubtedly, the implementation of PARS helped to reduce the training time in comparison to other state-of-the-art DRL algorithms for large-scale grid control problems. Still, upon investigation, it is found that if we can improve the computation time consumed by the power system simulator for the rollout generation, the training time can be reduced further to make it almost real-time.
\begin{figure}[t]
  \centering
    \includegraphics[width=0.48\textwidth]{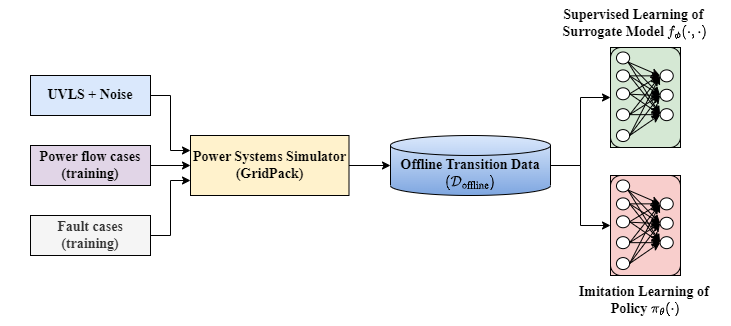}
  \caption{Initial Phase: Surrogate Model Training and Imitation Learning}
  \label{f1}
 \vspace*{-0.1in}
 \end{figure}
 \begin{figure}[t]
  \centering
    \includegraphics[width=0.48\textwidth]{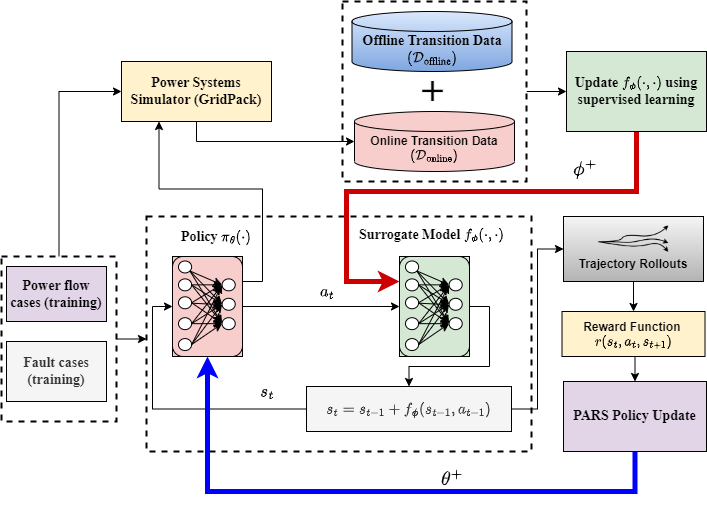}
  \caption{Flowchart of MB-PARS policy update}
  \label{f22}
 \vspace*{-0.1in}
 \end{figure}
The key steps of MB-PARS algorithm are shown in \textbf{Algorithm 1}. The conceptual architecture of \textbf{Algorithm 1} is shown in Fig.~\ref{f1} and \ref{f22}. It is clear that the learning process can broadly divided into following major parts: 
\subsubsection{Generation of Offline Training Data} We utilized some given rule-based policy, for instance under voltage load shedding (UVLS) rule, mixed with random noise to generate the system trajectories for $\mathcal{T}$ tasks. The offline data set $\mathcal{D}_{\text{offline}}$ is built by preprocessing the trajectory data as multi-step transition tuples.

\subsubsection{Learning of Surrogate Model} A DNN-based surrogate dynamic model $f_{\phi}(\cdot,\cdot)$, $\phi$ representing DNN parameters, is trained on $\mathcal{D}_{\text{offline}}$ minimizing the multi-step loss function given in (\ref{eqn_mult_step_loss}). We utilize the trained surrogate model (faster in computation speed) to generate the rollouts while exploring the parameter space of DRL policy at step-12.

\subsubsection{Imitation Learning-based Policy Initialization}
The basic idea behind imitation learning-based initialization is to provide a warm-start in the MB-PARS policy search. We used UVLS-based policy to collect training data set for surrogate model. Therefore, without collecting additional data, we conducted a supervised learning to learn an initial policy using stored offline data. Note that, this policy is not an optimal one, but somehow mimics the UVLS policy. We initialized our MB-PARS policy with the weights of this initial policy. Compared to using randomized initial weights for the MB-PARS policy, which is the common practice, imitation learning could greatly reduce the searching period at the beginning of policy optimization process.

\subsubsection{Policy Learning} The policy network $\pi(\theta)$ is modeled using an long short term memory (LSTM) network. LSTM is a type of recurrent neural network capable of learning long-term dependencies. It should be noted that the proposed MB-PARS is utilized for voltage control problems; that is why LSTM is used to learn the temporal correlation of the voltage observations. As mentioned earlier, policy learning is achieved through the derivative-free PARS algorithm. But, before starting the policy training, the weights $\theta$ of the policy network is initialized by leveraging imitation learning, as discussed earlier..

\subsubsection{Online Retraining of Surrogate Model} To mitigate the issue that arises from the distribution mismatch, as mentioned in Section \ref{sec2c}, the algorithm generates an online ground-truth data sample with the current policy. Then it retrains the surrogate model at a given frequency (step-17 to step-20) with a combined set of online and offline data.  

\begin{algorithm}[!htbp]
	\caption{MB-PARS} 
	\begin{algorithmic}[1]\label{algo1}
		\State Collect trajectory data offline for task set $\mathcal{T}$ using a UVLS policy mixed with random noise. Create multi-step transition tuples and store it in $\mathcal{D}_{\text{offline}}$. 
		\State Instantiate a DNN-based surrogate dynamic model $f_{\phi}(\cdot,\cdot)$, where $\phi:=$ weights and bias of DNN. Train $f_{\phi}(\cdot,\cdot)$ on $\mathcal{D}_{\text{offline}}$ minimizing the multi-step loss function given in (\ref{eqn_mult_step_loss}) and save the trained model $f_{\phi}(\cdot,\cdot)$ (Details of this supervised learning step can be found in Section \ref{sec2b}). Set surrogate model update frequency $UF$.
		\State Instantiate the LSTM-based policy $\pi(\theta)$. 
		\State Conduct imitation learning using dataset $\mathcal{D}_{\text{offline}}$. Initialize the policy weights $\theta_0 \in \mathbf{R}^{n\times p}$ with the imitation learning policy weights.

		\State Set the hyper-parameters required for policy training, $\alpha:=$ step size, $N:=$ number of policy perturbation direction per iteration, $v:=$ standard deviation of exploration noise, $b:=$ number of top-performing directions, $m:=$ number of rollouts per perturbation direction, and $\epsilon:=$ decay rate.
		\State Set $\mu_0 = \mathbf{0} \in \mathbf{R}^n$, the running mean of observation states ($s_t$), and $\Sigma_0 = \mathbf{I} \in \mathbf{R}^{n \times n}$, the running standard deviation of observation states ($s_t$), total iteration number $H$.
	    \For {iteration $k=1,H$}
	    \State Randomly select $N$ directions $\delta_1,\cdots,\delta_N \in \mathbf{R}^{n\times p}$ \newline
        \hspace*{1em} (dimension same as of policy $\theta$) for policy perturbation.
        \State Call the current version of surrogate model $f_{\phi}(\cdot,\cdot)$.
	    \For {each $\delta_i \vert_{i=0}^{N}$}
	    \State Perturb the policy weights $\theta$ in both $\pm$ direction \newline
        \hspace*{2.5em} of $\delta_i$: $\theta_{ki+}=\theta_{k-1}+v\delta_i$ and $\theta_{ki-}=\theta_{k-1}-v\delta_i$
	    \State Sample $p$ tasks randomly from a task set $\mathcal{T}$, and \newline 
	    %\For {each task $1\rightarrow p$}
        \hspace*{2.8em} for each task $p$, observe the initial state $s_0$, and \newline 
        \hspace*{2.8em} generate total $2\times m$ rollouts or episodes using the \newline 
        \hspace*{2.8em} $f_{\phi}(s_{k,t},a_{k,t})$.
        \State During each rollout, normalize the states $s_{k,t}$ \newline
        \hspace*{2.8em} at time step $t$, using $s_{k,t} = \frac{(s_{k,t} - \mu_{k-1})}{\Sigma_{k-1}}$; next \newline
        \hspace*{2.8em} obtain the action $a_{k,t} = \pi(\theta_k,s_{k,t})$, and get the \newline
        \hspace*{2.8em} new state using $s_{k,t+1} = s_{k,t} + f_{\phi}(s_{k,t},a_{k,t})$. \newline
        \hspace*{2.8em} Update $\mu_k$ and $\Sigma_k$ with $s_{k,t+1}$. 
	    \State Calculate the average rewards over $m$ rollouts\newline
        \hspace*{2.8em} $\{\mathbf{r}_{ki+},\mathbf{r}_{ki-}\}$, respectively for $\pm$ perturbation. 
	    \EndFor
	    \State Select the top $b$ directions among $\delta_1,\cdots,\delta_N$ based on \newline 
        \hspace*{1em} the $\max{\{\mathbf{r}_{ki+},\mathbf{r}_{ki-}\}}$ and calculate their standard \newline 
        \hspace*{1em} deviation $\sigma_b$.
        \State $\theta_{k+1} = \theta_k + \frac{\alpha}{b\sigma_b}\sum_{i=1}^{b}(\mathbf{r}_{ki+}-\mathbf{r}_{ki-}) \rightarrow$ Update policy \newline 
        \hspace*{1em} weight.
        % \begin{align}
        %     \theta_{k+1} = \theta_k + \frac{\alpha}{b\sigma_b}\sum_{i=1}^{b}(r_{ti+}-r_{ti-})
        % \end{align}
        \State Evaluate the current policy $\theta_{k+1}$ in the ground-truth \newline 
        \hspace*{1em} environment and generate new data set with current \newline 
        \hspace*{1em} policy for all tasks $\in \mathcal{T}$. Create new multi-step \newline 
        \hspace*{1em} transition tuples and store it in $\mathcal{D}_{\text{online}}$.
        \If {${k}/{UF}$ = 0 \textbf{and} $k > 0$}
        \State Retrain the model $f_{\phi}(\cdot,\cdot)$ with a combined data \newline 
        \hspace*{2.5em} set $\mathcal{D} = \mathcal{D}_{\text{online}} + 25 \% \;\text{of}\; \mathcal{D}_{\text{offline}}$ using supervised \newline 
        \hspace*{2.5em} learning.
        \State \textbf{return} Updated surrogate model $f_{\phi}(\cdot,\cdot)$.
        \EndIf 
        \State Decay $\alpha$ and $v$ with rate $\epsilon$: $\alpha=\epsilon \alpha$, $v = \epsilon v$.
		\EndFor
		\State \textbf{return} $\theta$
	\end{algorithmic} 

\end{algorithm}

\subsection{Emergency Voltage Control: An RL Formulation}
\begin{figure}[t]
  \centering
    \includegraphics[width=0.4\textwidth]{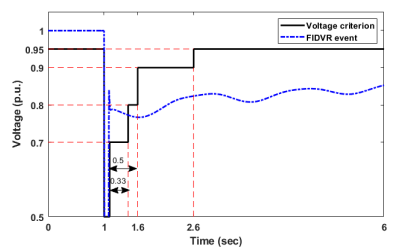}
  \caption{Transient Voltage Recovery Criterion \cite{pjm}}
  \label{f2}
 %\vspace*{-0.2in}
 \end{figure}

This paper considers a short-term voltage instability problem originating from Fault-induced delayed voltage recovery (FIDVR). FIDVR has occurred in various US utilities. Briefly, it is defined as the phenomenon whereby system voltage remains at significantly reduced levels for several seconds after a fault has been cleared. In general, stalling of air-conditioner (A/C) motors (1-phase induction motors) and prolonged tripping is the root cause of FIDVR. To mitigate FIDVR, it is required to have a well-proof emergency control plan according to the voltage recovery criterion given in \cite{pjm} (shown in Fig.~\ref{f2}). Traditionally, rule-based under-voltage load shedding (UVLS) schemes are used as a part of the control plan \cite{taylor1992concepts}. But, in general, UVLS policy does not provide an optimal solution and may cause unnecessary tripping of essential loads. To find the optimal solutions, we need to formulate the problem as a non-convex, nonlinear constrained optimization as detailed in our recent work \cite{huang2019adaptive}. Reference \cite{huang2019adaptive} also provides the details of the MDP formulation of the above mentioned optimization problem to implement RL-based control schemes. Next, we briefly discuss the MDP structure of the load shedding based voltage control problem for completeness. 
\subsubsection{State} In voltage control problem with load shedding as control action, obviously the main variables of interest are, $V_t := [V^1_t,\cdots,V^m_t]^'$: the voltage measurement of monitored buses at time $t$ and $P_{Dt} := [P_{Dt}^1,\cdots,P_{Dt}^n]^'$ the percentage load remaining at the controlled buses at time $t$, where $m$ and $n$ are respectively no. of monitored bus and no. of controlled bus. Hence, we stacked them together to form $s_t := [V^1_t,\cdots,V^m_t,P_{Dt}^1,\cdots,P_{Dt}^n]^'$ the state of the underlying MDP.
\subsubsection{Action} The action $a_t$ is to perform load shedding at each of the $n$ no. of controlled buses. We design the load shedding as a continuous control action, and at each time step it can vary from $0$ to $20\%$ of the respective bus load. Thus, $a_t = [a_t^0 \cdots a^n_t]^'$, and the action space is $[-0.2,0]$ (minus means shedding).
\subsubsection{State Transition} As mentioned earlier, in this paper state transition is achieved through the learned surrogate model using the relation: $s_{t+1} = s_{t} + f_{\phi}(s_{t},a_{t})$.
\subsubsection{Reward} The training of the agent depends solely on the reward structure. We follow the reward structure used in our previous work \cite{huang2021accelerated,huang2019adaptive} with certain vital modifications. The main objective of this problem is to meet the voltage recovery criteria as shown in Fig.~\ref{f2} with a minimum amount of load shedding. According to the standard, the voltages should return to at least 0.8, 0.9, and 0.95 p.u. within 0.33, 0.5, and 1.5 s. Overall the design principle is as follows: if the agent fails to recover the voltage above 0.95 p.u. even after 4 sec. of $T_{pf}:=$ the time instant of fault clearance, we penalize the agent heavily with a large negative number $-R$, while in case of success of recovering the voltages above 0.95 p.u. after $T_{pf}+4$ requires to compute the reward as a weighted sum of voltage deviations (according to the specified standard), the amount of load shedding (voltage recovery with minimum possible load interruption) and the penalty for invalid action (action of load shedding even when the load at the bus is 0). The reward function used in \cite{huang2021accelerated} addresses all the issues mentioned above, but in context of the current problem where we are utilizing the surrogate model to generate rollouts, we need to tackle some other issues that originate from the inaccuracy of the surrogate model dynamics. Thus, we introduced two major changes in the reward equation compared to the one used in \cite{huang2021accelerated}. The reward $r_t$ is shown in (\ref{eqnrew1})-(\ref{eqnrew2}).

\small
\begin{equation}\label{eqnrew1}
r_t =
\begin{cases}
   -R, \;\;\;\;\Scale[0.9]{\text{if} \;V^i_t< (0.95-d) \;\;\text{and}\;\;  t > T_{pf} + 4} \\
    -R \times e^{-[\min_{i}\{V^i_t\} - (0.95-d)]\times\tau},\\\;\;\;\;\;\;\;\;\;\;\;\;\;\;\;\;\Scale[0.9]{\text{if} \;(0.95-d)\leq V^i_t<0.95\;\;\text{and}\;\;t > T_{pf} + 4} \\
   c_1\sum_{i}{\Delta{V^i_t}} - c_2\sum_{j}\Delta{P^j_{t}}-c_3u_{ilvd},\; \Scale[0.9]{\text{otherwise}}
\end{cases}
\end{equation}
\begin{equation}\label{eqnrew2}
\Delta V^i_t = 
\begin{cases}
  \min\{V^i_t-0.7,0\} \;\;\; \Scale[0.9] {\text{if} \; T_{pf} <t<T_{pf} +0.33}\\
  \min\{V^i_t-0.8,0\} \;\;\; \Scale[0.9]{\text{if} \; T_{pf}+0.33 <t<T_{pf} +0.5}\\
  \min\{V^i_t-0.9,0\} \;\;\; \Scale[0.9]{\text{if} \; T_{pf} +0.5<t<T_{pf} +1.5}\\
  \min\{V^i_t-0.95,0\} \;\;\; \Scale[0.9]{\text{if} \; T_{pf}+1.5 <t}
\end{cases}
% \vspace*{-.05in}
\end{equation}
\normalsize
where $V^i_t:=$  is the bus voltage magnitude for bus $i$ at time step $t$, $\Delta{P^j_{t}}$ is the load shedding amount in p.u. for load bus $j$ at time step $t$, $u_{ilvd}$ is the invalid action penalty, $c_1, c_2, c_3$ are weights of the three components used in computing the reward when the agent meet the minimum voltage recovery criteria. 

To tackle the model inaccuracy and maintain learning stability, we introduced a soft penalty if the agent fails to meet the minimum voltage recovery criteria. In other words, instead of penalizing the agent heavily if the voltage computed by the surrogate model is close but less than 0.95 p.u. after $T_{pf}+4$ (minimum voltage recovery criteria), we penalize the agent according to the soft function $-R \times e^{-[\min_{i}\{V^i_t\} - (0.95-d)]\times\tau}$. Here, we used $d$ as a dead-band near the minimum required voltage (i.e., 0.95 p.u.) and $\tau$ as a time constant. 

In earlier model-free PARS implementation \cite{huang2021accelerated}, the values of the weights $c_1$ and $c_2$ were such that there is an implicit jump in the reward function near minimum voltage recovery criteria. This jump notoriously affected the learning of the agent in the current surrogate model-based framework. But, this jump does not have any effect while learning through the actual simulator. This made us modify the values $c_1$ and $c_2$ to eliminate such implicit jump to complete the training successfully even in the presence of inaccuracy originated from the use of the surrogate model.

%\subsection{DNN based Surrogate Model for voltage control}

% \subsection{Imitation Learning}
% Imitation learning has achieved great successes in recent years in the machine learning field \cite{10.1145/3054912}. This technique aims to mimic an expert policy before the reinforcement learning training stage to give the policy a warm start, which usually reduces training time in the early searching stage.

% The basic idea behind imitation learning is to let the policy learn from an existing easy-to-obtain policy before training it through interactions with the environment. The existing policy chosen in this study is the UVLS policy. The same offline dataset which was generated in the surrogate model training stage were utilized for this procedure.

% By conducting supervised learning, initial weights of a LSTM policy that is similar to the UVLS policy were obtained and used as the starting weights in the DRL training. Compared to using randomized initial weights for the DRL policy, which is the common practice, imitation learning could greatly reduce the beginning searching period of the policy optimization process.

%\subsection{ARS algorithm}

% \subsection{Design of DNN based Surrogate Model for voltage control}

% 

% \subsection{ARS algorithm}

% \subsection{MBRL steps}

% \subsection{Architecture Figure and explanation}

% \subsection{Details of the reward design}

%\input{4_Imitation_Learning}

\section{Test Cases and Results} \label{sec4}

In this section, we first describe the training and testing environment for implementing the proposed method. Then we present the simulation results and comparison with model-free PARS to demonstrate the superiority of the proposed methods in terms of training efficiency and control performance.%, and RL agent generalization capacity.

\subsection{Test case: IEEE 300-Bus System}

The training and testing of the proposed method are performed with a modified IEEE 300-bus system with loads larger than 50 MW within Zone 1 represented by WECC composite load model {\cite{huang2019_cmpldw}}. The power system dynamic simulation is completed by the open-source platform GridPACK \cite{huang2017faster}. 

In the training phase, 24 operational scenarios comprising 4 power flow cases and 6 fault buses are considered. The faults are applied at 1.0 s at the respective buses, and self cleared after a duration of 0.1 sec. While, in the testing phase, we considered 132 operational scenarios comprising 4 power flow scenarios and 33 fault buses. Compared with the training cases, we added 27 more fault buses during testing. The reason for applying new test scenarios is to validate the adaptability of the proposed data-driven control policy. The detailed power flow conditions for training and testing are shown in Table~\ref{tab_pfcases} and \ref{tab_faultlocation} show the details of the power flow and fault cases, respectively. 

% The training and testing of the proposed method are performed with a modified IEEE 300-bus system with loads larger than 50 MW within Zone 1 represented by WECC composite load model {\cite{huang2019_cmpldw}}. The power system dynamic simulation is completed by the open-source platform GridPACK \cite{huang2017faster}. For training the algorithm, 24 operation scenarios are generated, which combines 4 power flow scenarios with 6 fault scenarios. The power flow scenarios vary in their generation levels and loading levels, and the fault takes places at 6 different buses. The fault is assumed to start at 1.0 s and ends after 0.1 s. For testing the algorithm, 132 operation scenarios are generated, which combines 4 power flow scenarios with 33 fault scenarios. Compared with the training cases, 27 more fault buses are considered during testing. The reason for applying new test scenarios is to validate the adaptability of the proposed data-driven control policy. The detailed power flow conditions for training and testing are shown in Table~\ref{tab_pfcases}. The fault locations for training and testing are shown in Table~\ref{tab_faultlocation}.

\begin{table}[t]
\caption{Power flow scenarios for training and testing}
\tabcolsep=0.045 cm
\begin{center}
\begin{tabular}{|c|c|c|}
\hline
\textbf{Power flow scenarios} & \textbf{Generation} & \textbf{Load} \\
\hline
{\textbf{Scenario 1}} & \makecell[c]{100\% for all generation\\(21372.6 MW)}&\makecell[c]{100\% for all loads\\(20950.1 MW)}\\ 
\hline
{\textbf{Scenario 2}}& 115\% for all generators & 115\% for all loads \\ 
\hline
{\textbf{Scenario 3}}& 85\% for all generators & 85\% for all loads \\ 
\hline
{\textbf{Scenario 4}}& 92\% for all generators & 80\% for loads in Zone 1 \\ 
\hline
\end{tabular}
\label{tab_pfcases}
\end{center}
\end{table}

\begin{table}[t]
\caption{Bus indices of fault locations for training and testing}
\begin{center}
\tabcolsep=0.2 cm
%\begin{tabular}{|p{3cm}|p{3cm}|}
\begin{tabular}{|c|c|}
\hline
\textbf{Training} & \textbf{Testing}\\
\hline
{3,5,8,12,17,23}&{3,5,8,12,17,23,}\\
{}&{26,1,4,6,7,9,10,11,13,}\\
{}&{14,16,19,20,21,22,25,87,102,}\\
{}&{89,125,160,320,150,123,131,130}\\
\hline
\end{tabular}
\label{tab_faultlocation}
\end{center}
\end{table}

Load-shedding control actions are considered for all buses with a dynamic composite load model at Zone 1 (34 buses in total). The load shedding percentage at each control step can vary from 0 to 20\%. The agent is designed to provide control decisions (including no action as an option) to the grid at every 0.1 seconds.
The observations included voltage magnitudes at 142 buses within Zone 1 and the remaining fractions of 34 composite loads. We also included the fault information as part of the input values for the policy training, consisting of the fault bus, fault start time, fault duration, and the time distance of the current step to the fault start time. Thus the dimension of the observation space is 180. 
%The agent needs to obtain the observations from the grid environment every 0.1 seconds.
The PARS algorithm without a surrogate model is the baseline method we used to compare the proposed model-based approach. The goal is to achieve similar control performance while greatly reducing the training time of the reinforcement learning algorithm.
% Load shedding control actions are considered for all buses with dynamic composite load model at Zone 1 (34 buses in total). The percentage of load shedding at each control step could be from 0 to 20\%. The agent is designed to provide control decisions (including no action as an option) to the grid every 0.1 second.
% The observations included voltage magnitudes at 142 buses within Zone 1, as well as the remaining fractions of 34 composite loads. We also included the fault information as part of the input values for the policy training, which was consist of the fault bus, fault start time, fault duration, and the time distance of current step to the fault start time, thus the dimension of the observation space is 180. The agent needs to obtain the observations from the grid environment every 0.1 second.

% The baseline method we used to compare the proposed model-based approach is the PARS algorithm without a surrogate model. The goal is to achieve similar control performance while greatly reducing the training time of the reinforcement learning algorithm.  

\begin{table}[t]
\caption{Hyperparameters for training 300-bus system}
\begin{center}
\begin{tabular}{|l|c|c|}
\hline
\textbf{Parameters} & \textbf{300-Bus} \\
\comment{
\hline
Policy Model & Linear, FNN, LSTM \\}
\hline
Policy Network Size (Hidden Layers) & [32,32]  \\
\hline
Number of Directions ($N$) & 60  \\
\hline
Top Directions ($b$) & 30  \\
\comment{
\hline
Number of Fault Cases ($F$)& 6\\}
\comment{
\hline
No. of Maximum Iterations ($I$) & 300\\}
\comment {
\hline
Rollout Length & 80 \\}
\hline
Step Size ($\alpha$) & 1 \\
\hline
Step Size ($\alpha$) (with IL) & 0.05 \\
\hline
Std. Dev. of Exploration Noise ($\upsilon$) & 2  \\
\hline
Std. Dev. of Exploration Noise ($\upsilon$) (with IL)  & 0.1  \\
\hline
Decay Rate ($\varepsilon$) & 0.9985  \\
\hline
Decay Rate ($\varepsilon$) (with IL) & 0.9999  \\
\hline
\end{tabular}
\label{tab_IEEE300}
\end{center}
\end{table}

\subsection{Surrogate model training}
\begin{figure}[t]
  \centering
    \includegraphics[width=0.48\textwidth]{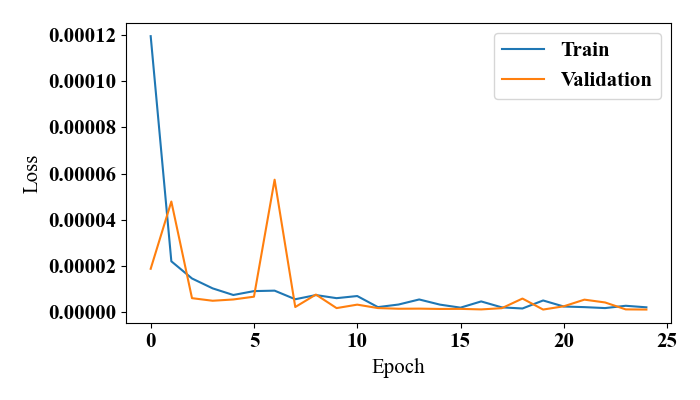}
  \caption{Losses of the Surrogate Model}
  \label{model_loss}
 %\vspace*{-0.2in}
 \end{figure}

A fully-connected neural network (FCNN) is considered to represent the surrogate model $f_{\phi}(\cdot,\cdot)$, which replicates the power system dynamics. The FCNN has 3 hidden layers of size $[1000, 500, 200]$ with ReLU as the nonlinear activation function. The activation function considered for the output layer is sigmoid. For the training of the surrogate model, we created an offline data set utilizing (a) a UVLS policy and adding random noise to the obtained actions and (b) power flow and fault scenarios defined under the training phase. With this offline data set, we conducted supervised learning to train the surrogate model. Note the training process follows the steps in Section II.B to create the multi-step loss function defined in (\ref{eqn_mult_step_loss}). In this experiment, we choose a 5-step loss, i.e., $M=5$. The plots of training and validation losses of the surrogate model are shown in Fig.~\ref{model_loss}, which indicates a good convergence.  

\subsection{Policy training}
\begin{figure}[t]
  \centering
    \includegraphics[width=0.48\textwidth]{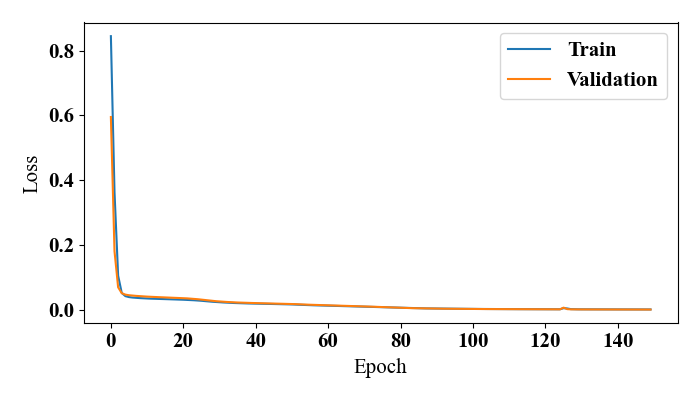}
  \caption{Losses of the Imitation Learning Policy}
  \label{IL_loss}
 %\vspace*{-0.2in}
 \end{figure}

First, we conducted imitation learning using the same offline data collected for surrogate model learning. Imitation learning initializes the weights of the policy network, which is an LSTM network for this study. Fig.~\ref{IL_loss} shows a good convergence of training and validation loss during the training of the imitation learning policy. 

With the trained surrogate model, and imitation learning-based initialized policy network, we started MB-PARS policy training following Algorithm 1. The hyperparameters used in the training of MB-PARS policy are shown in Table ~\ref{tab_IEEE300}. Fig.~\ref{Training} shows the comparison of average rewards during the RL policy training for the PARS algorithm using the model-based approach (using surrogate model, labeled as `Model-based' in Fig.~\ref{Training}) and model-free approach (using GridPACK simulator, labeled as `PARS' in Fig.~\ref{Training}) with or without imitation learning. Results were obtained by averaging over five different random seeds. In Fig.~\ref{Training}, the $x$-axis presents the actual clock time; hence, the plot compares performance with respect to actual clock time. There are some important points, as mentioned below, to note regarding this result.
\begin{itemize}
    \item The starting time of the model-based approach is later than the model-free counterpart. This is done to add the offline training time for the surrogate model, which is taken into consideration to have a fair comparison regarding algorithm efficiency.
    % \item The comparison of MB and MB + IL shows the impor
    % \item \textcolor{blue}{There is also a difference between MB and MB + IL that a simpler surrogate model is sufficient for MB + IL to converge which leads to reduced offline training time.}
    \item Different starting reward values in Fig.~\ref{Training} clearly shows that imitation learning provides a better start in policy search than starting from a randomized initial policy. 
    \item Imitation learning reduces the variance in average reward during training, and helps the training process to stabilize in case of model-based approach, where the training performance is also dependent on the learned surrogate model. Our investigation also finds that use of imitation learning reduces the chance of training divergence. 
\end{itemize}

% First is that the model-based approach starting time is later than using the GridPACK. This is due to the offline training time for the surrogate model, which was taken into consideration to have a fair comparison regarding algorithm efficiency. \textcolor{blue}{There is also a difference between MB and MB + IL that a simpler surrogate model is sufficient for MB + IL to converge which leads to reduced offline training time.} Another thing to note is that by using imitation learning, the starting policies of these \textcolor{blue}{four} approaches are different. The imitation learning starting policy is much better than the randomized initial policy of PARS, which can be seen by the starting average rewards of the \textcolor{blue}{four} approaches. \textcolor{blue}{Another important point which can be seen here is} that imitation learning reduced the variance during training. \textcolor{blue}{This is important because IL stabilizes the MB approach and reduces the chance of training not converging. In our study we discover IL to be a way for boosting the MB training.}

%When only comparing one iteration time in training, model-based vs. original PARS algorithm is around 15s vs. 150s. However, because of the gap between surrogate model and GridPACK, model-based training generally needs more iterations to converge. 
\subsection{Comparison of training time and sample efficiency}
Our proposed MB-PARS approach with IL (MB +IL) converged around 800s, while the state-of-the-art model-free PARS algorithm \cite{huang2021accelerated} converged around 6500s. This clearly shows an 87.7\% reduction in training time. It is also important to note that rollout generation in model-free PARS is achieved with GridPACK, one of the fastest simulators in power system. Additionally, compared to PARS + IL, our proposed method (MB + IL) started late (due to offline surrogate model training) but catches up quickly and converges faster than the PARS + IL approach.

\begin{figure}[t]
  \centering
    \includegraphics[width=0.48\textwidth]{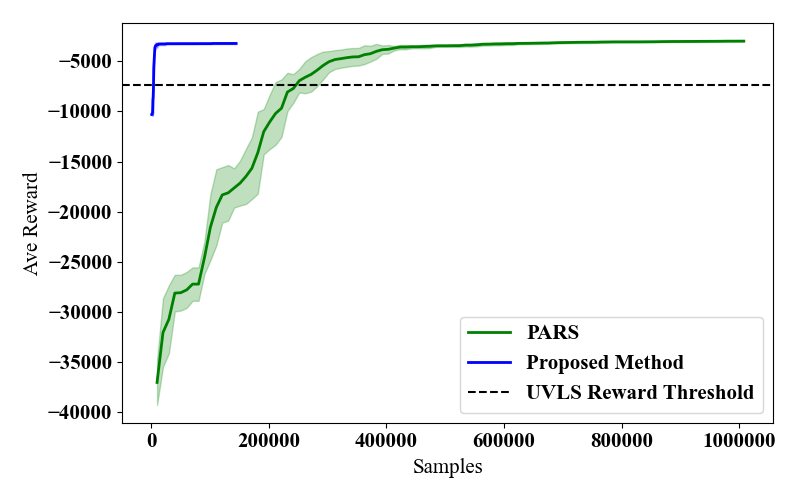}
  \caption{MB + IL Policy Training Sample Efficiency Comparison}
  \label{Training_sample}
 %\vspace*{-0.2in}
 \end{figure}

Next, we compare the sample efficiency. Fig. \ref{Training_sample} shows the comparison of sample efficiency between the baseline method PARS and our proposed method MB + IL. The MB + IL approach converged around 15000 samples, while the PARS algorithm converged around 600000 samples. This shows a 97.5\% reduction in samples needed from the power system simulator.

%Figure shows both the model-based and GridPACK-based approaches training results with the help of imitation learning. It can be seen that both algorithms have a much better starting rewards compared to without imitation learning. This helps the algorithm to converge in a much faster way. MB plus IL converged in less than 1000s, while PARS + IL converged at around 2000s. One more thing to note is that with IL, the surrogate model needed can be simplified too, thus the shorter offline training time compared to without imitation learning.
%When comparing all four approaches in Figure, we can conclude with two key take-aways: 1). PARS with model-based training is faster than without model-based. 2). PARS with imitation learning is faster than without it.

%When utilizing both the surrogate model and imitation learning, the training time reduced from more than 6000s to less than 1000s.


\begin{figure}[t]
  \centering
    \includegraphics[width=0.48\textwidth]{figure/four_methods_newfont.png}
  \caption{MB + IL Policy Training Average Rewards Comparison}
  \label{Training}
 %\vspace*{-0.2in}
 \end{figure}

\subsection{Policy testing}

To show the adaptability of the trained policy in unknown scenarios, we tested our trained policy with 33 different fault buses with 5 different power flow scenarios (1 training power flow scenarios and 4 testing power flow scenarios). Table ~\ref{tab_pfcases} shows the testing power flow scenarios. The testing power flow, and fault cases are given in Table~\ref{tab_pfcases} and  \ref{tab_faultlocation}, respectively.

\begin{figure}[t]
  \centering
    \includegraphics[width=0.48\textwidth]{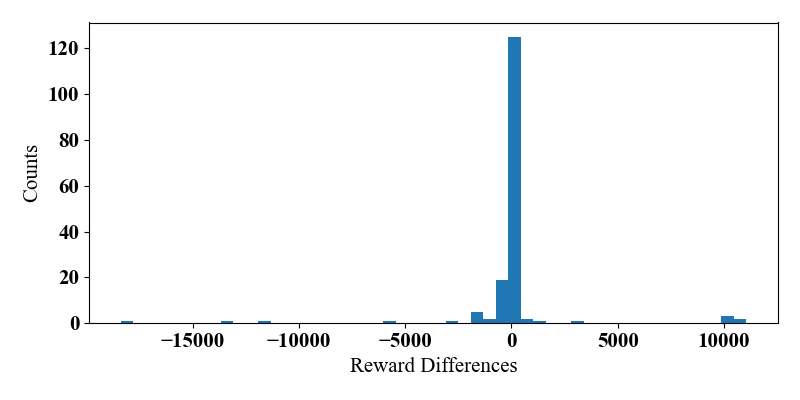}
  \caption{Reward differences between proposed method to PARS}
  \label{rew_diff}
 %\vspace*{-0.2in}
 \end{figure}

The reward differences between the model-based policy and the PARS policy are shown in Fig. ~\ref{rew_diff}. It can be seen that the two policies are very similar in most testing cases while either policy is better in a few cases. Regarding the average reward, PARS policy is slightly better, achieving -3828.91 compared to -3859.60 by our proposed method. Table ~\ref{comparison_table} shows the comparison results.

\begin{table}[t]
\caption{Comparisons Between Proposed Method and PARS}
\begin{center}
\begin{tabular}{| c| c| c| }
\hline
  & \multicolumn{2}{|c|}{Method} \\ 
\hline
 Metrics & Proposed & PARS \\ 
\hline
 Convergence Time & 800s & 6500s \\  
\hline
Convergence Samples & 15000 & 600000 \\  
\hline
 Testing Reward & -3859.60 & -3828.91 \\
 \hline
\end{tabular}
\label{comparison_table}
\end{center}
\end{table}

To show the voltage recovery with trained policies, we plotted the voltage curves and corresponding load-shedding for fault at bus 13. In this particular case, even though both methods achieved the desired voltage performance, but our proposed method performed better than the PARS policy in terms of the amount of load-shedding.

% Fig. \ref{fault21} is a single testing case example where the proposed method performed better than PARS. This is the voltage recovery for bus 13 for a fault at bus 21. Fig. \ref{fault21_load} shows the total load shed of the two methods. 

%We can see that the proposed method achieved similar recovery result with PARS.

\begin{figure}[htbp!]
  \centering
    \includegraphics[width=0.48\textwidth]{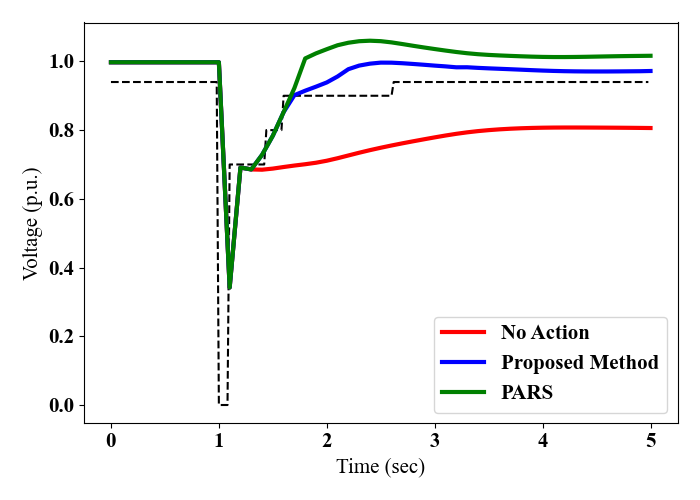}
  \caption{Plot of voltages at Bus 13 for fault at Bus 21}
  \label{fault21}
 \vspace*{-0.2in}
 \end{figure}

\begin{figure}[htbp!]
  \centering
    \includegraphics[width=0.48\textwidth]{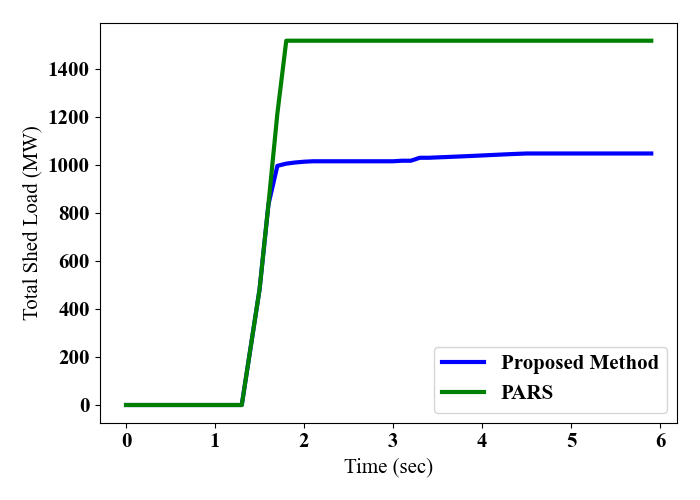}
  \caption{Total Load Shed for fault at Bus 21}
  \label{fault21_load}
 \vspace*{-0.2in}
 \end{figure}
%\subsection{Multi-step loss training}
%\subsection{Model update during RL training}

\section{Conclusions}\label{sec:conclusions}
This paper proposes a model-based training approach for the PARS algorithm, aided by imitation learning, as a solution to the FIDVR problem. The proposed algorithm utilized a surrogate model that learned about the power system dynamics to generate roll-outs in the RL training stage to reduce the training time. We also added imitation learning to this process to provide a warm start to the RL policy, thus reducing the early searching time of the policy training. By testing the proposed approach in the IEEE 300-bus system, we show that the policy trained using the surrogate model with imitation learning achieved similar control performance with the PARS policy trained using GridPACK while needing only 13\% of the training time as PARS. This result is significant because long training time is the bottleneck for many RL applications, especially in the power system research field. With the fast-changing nature of the power systems, an efficient training procedure requiring only one-tenth of the training time of the previous method provides a promising avenue for future research and applications.

% For future research, ...

% if have a single appendix:
%\appendix[Proof of the Zonklar Equations]
% or
%\appendix  % for no appendix heading
% do not use \section anymore after \appendix, only \section*
% is possibly needed

% use appendices with more than one appendix
% then use \section to start each appendix
% you must declare a \section before using any
% \subsection or using \label (\appendices by itself
% starts a section numbered zero.)

\comment{
\appendices
\section{Proof of the First Zonklar Equation}
Appendix one text goes here.

% you can choose not to have a title for an appendix
% if you want by leaving the argument blank
\section{}
Appendix two text goes here.

% use section* for acknowledgment
\section*{Acknowledgment}

The authors would like to thank xxx from U.S. DOE ARPA-E for his support and guidance.

}

% Can use something like this to put references on a page
% by themselves when using endfloat and the captionsoff option.
\ifCLASSOPTIONcaptionsoff
  \newpage
\fi

% trigger a \newpage just before the given reference
% number - used to balance the columns on the last page
% adjust value as needed - may need to be readjusted if
% the document is modified later
%\IEEEtriggeratref{8}
% The "triggered" command can be changed if desired:
%\IEEEtriggercmd{\enlargethispage{-5in}}

% references section

% can use a bibliography generated by BibTeX as a .bbl file
% BibTeX documentation can be easily obtained at:
% http://mirror.ctan.org/biblio/bibtex/contrib/doc/
% The IEEEtran BibTeX style support page is at:
% http://www.michaelshell.org/tex/ieeetran/bibtex/
%\bibliographystyle{IEEEtran}
% argument is your BibTeX string definitions and bibliography database(s)
%\bibliography{IEEEabrv,../bib/paper}
%
% <OR> manually copy in the resultant .bbl file
% set second argument of \begin to the number of references
% (used to reserve space for the reference number labels box)

\bibliographystyle{IEEEtran}
\bibliography{mbrl.bib}

% biography section
% 
% If you have an EPS/PDF photo (graphicx package needed) extra braces are
% needed around the contents of the optional argument to biography to prevent
% the LaTeX parser from getting confused when it sees the complicated
% \includegraphics command within an optional argument. (You could create
% your own custom macro containing the \includegraphics command to make things
% simpler here.)
%\begin{IEEEbiography}[{\includegraphics[width=1in,height=1.25in,clip,keepaspectratio]{mshell}}]{Michael Shell}
% or if you just want to reserve a space for a photo:
\comment{
\begin{IEEEbiography}{Michael Shell}
Biography text here.
\end{IEEEbiography}

% if you will not have a photo at all:
\begin{IEEEbiographynophoto}{John Doe}
Biography text here.
\end{IEEEbiographynophoto}

% insert where needed to balance the two columns on the last page with
% biographies
%\newpage

\begin{IEEEbiographynophoto}{Jane Doe}
Biography text here.
\end{IEEEbiographynophoto}

% You can push biographies down or up by placing
% a \vfill before or after them. The appropriate
% use of \vfill depends on what kind of text is
% on the last page and whether or not the columns
% are being equalized.

%\vfill

% Can be used to pull up biographies so that the bottom of the last one
% is flush with the other column.
%\enlargethispage{-5in}

}

% that's all folks
\end{document}